\documentclass[sigconf]{acmart}

\newcommand{\jshelter}[0]{JShelter}
\newcommand{\NSCL}[0]{NSCL}
\newcommand{\NSCLfull}[0]{NoScript Commons Library}
\newcommand{\fsf}[0]{Free Software Foundation}
\newcommand{\noscript}[0]{NoScript Security Suite}

\usepackage[utf8]{inputenc}
\usepackage{url}

\setcopyright{none}
\copyrightyear{2023}
\acmYear{2023}
\acmDOI{XXXXXXX.XXXXXXX}
\acmISBN{}
\acmConference[]{}{}{}
\begin{document}

  \author{Libor Polčák}
  \orcid{0000-0001-9177-3073}
  \email{polcak@fit.vut.cz}
  \affiliation{
    \department{Faculty of Information Technology}
    \institution{Brno University of Technology}
    \streetaddress{Božetěchova 2}
    \postcode{612 66}
    \city{Brno}
    \country{Czech Republic}
  }

  \author{Marek Saloň}
  \affiliation{
    \department{Faculty of Information Technology}
    \institution{Brno University of Technology}
    \streetaddress{Božetěchova 2}
    \postcode{612 66}
    \city{Brno}
    \country{Czech Republic}
  }

  \author{Giorgio Maone}
  \email{giorgio@maone.net}
  \affiliation{
    \institution{Hackademix}
    \streetaddress{via Mario Rapisardi 53}
    \postcode{90144}
    \city{Palermo}
    \country{Italy}
  }

  \author{Radek Hranický}
  \orcid{0000-0001-6315-8137}
  \email{ihranicky@fit.vut.cz}
  \affiliation{
    \department{Faculty of Information Technology}
    \institution{Brno University of Technology}
    \streetaddress{Božetěchova 2}
    \postcode{612 66}
    \city{Brno}
    \country{Czech Republic}
  }

  \author{Michael McMahon}
  \email{michael@fsf.org}
  \affiliation{
    \position{Web developer}
    \institution{Free Software Foundation}
    \streetaddress{51 Franklin Street Fifth Floor}
    \city{Boston}
    \postcode{MA 02110}
    \country{USA}
  }

  \title{\huge \jshelter{}: Give Me My Browser Back}

  \begin{abstract}
The web is used daily by billions. Even so, users are not protected from many
    threats by default. This paper builds on previous web privacy and
    security research and introduces \jshelter{}, a webextension that fights to return
    the browser to users. Moreover, we introduce a
    library helping with common webextension development tasks and fixing
    loopholes misused by previous research. \jshelter{}
    focuses on fingerprinting prevention, limitations of rich web APIs,
    prevention of attacks connected to timing, and learning information about
    the device, the browser, the user, and the surrounding physical environment and
    location. During the research of sensor APIs, we discovered a loophole in the sensor timestamps that lets any
    page observe the device boot time if sensor APIs are enabled in Chromium-based
    browsers. \jshelter{} provides a fingerprinting report and other feedback
    that can be used by future web privacy research.
    Thousands of users around the world use the webextension every day.
\end{abstract}
  \keywords{Browser fingerprinting, web privacy, web security, webextension
  APIs, JavaScript}

\maketitle
\section{Introduction}

Most people interact with web pages daily. Nowadays, many
activities are often carried out exclusively in a web browser, including shopping, searching for
travel information, and performing leisure
activities such as gaming, business, and office work. For several years, browser vendors have been adding new
JavaScript APIs to solicit the development of rich web applications~\cite{Snyder_features}.

Web visitors are subject to hostile
tracking~\cite{inria_cookie_banners,ico_adtech,session-replay,InputEventsAgeGender,belgianDPATCF},
fingerprinting~\cite{fingerprinting_survey,fingerprint_fingerprinters}, and
malware~\cite{primeprobe,forcepoint}.
Some of the recently added APIs influence the privacy of the users. For example,
the Geolocation
API\footnote{\url{https://developer.mozilla.org/en-US/docs/Web/API/Geolocation}}
is beneficial for navigation in the real world.
However, some users are only
willing to share the location with some visited sites. In the case
of Geolocation API, browsers ask users for permission, but not all APIs need
user permission. Users cannot limit the precision of the Geolocation API.
However, occasionally they want to
share a more precise location (e.g. during navigation). Other times they want
to share the location with limited precision (e.g. they are exploring a
location unrelated to their current position).

This paper presents \jshelter{}, a web browser extension (webextension) that allows
users to tweak the browser APIs. Additionally, \jshelter{} detects and prevents
fingerprinting. Moreover, \jshelter{} blocks attempts to misuse the browser as a proxy to
access the local network. \jshelter{} educates users by explaining fingerprinting APIs in a
report. \jshelter{} integrates several previous research projects like Chrome
Zero~\cite{js0} and little-lies-based fingerprinting prevention
\cite{PriVaricator,FPRandom}. As current webextension APIs lack a
reliable way to modify JavaScript APIs in different contexts like iframes and
web workers, we needed to solve the reliable injection.
This paper introduces \NSCLfull{}
(\NSCL{})\footnote{\url{https://noscript.net/commons-library}} that
other privacy- and security-related webextensions can reuse to solve common
tasks like the reliable injection of JavaScript code into the page JavaScript
context before the page scripts can access the context.
We implemented \jshelter{} for Firefox
and Chromium-based browsers like Chrome, Opera, and Edge. So we gathered experience
from user feedback that can be valuable to other research projects.
This paper extends our SECRYPT paper~\cite{jshelter}\footnote{If you intend to
cite our work, please, prefer citing the original paper~\cite{jshelter}.}.

The evaluation shows that \jshelter{} prevents many attacks, including learning
(1) browser and device fingerprints, (2) user biometrics, (3) computer clock-skew, and
(4) running applications. Sensors available for all pages on Android make
users vulnerable to several attacks~\cite{accomplice,accelerometer_fingerprinting,walking_patterns}.
\jshelter{} prevents the danger
by pretending to be a stationary device. \jshelter{} mitigates leaking boot time of the
device through sensor timestamps in Chromium-based browsers.

This paper is organised as follows. Section~\ref{sec:threats} presents the
threats that users face while web browsing.
Section~\ref{sec:countermeasures} compares \jshelter{} to
other security- and privacy-related webextensions. Section~\ref{sec:methodology}
provides the design decisions that we faced during the development of
\jshelter{}.
Section~\ref{sec:results} evaluates the \jshelter{} features and discusses
user feedback.
Section~\ref{sec:conclusion} concludes this paper.

\section{Threats}
\label{sec:threats}

Web Technology
Surveys\footnote{\url{https://w3techs.com/technologies/history_overview/client_side_language/all/y}}
report that 97.8\% of pages contain JavaScript, and the percentage is rising
over time. As the code is
unknown to the user, it might work against user
expectations~\cite{ico_adtech,belgianDPATCF}.
This section presents threats that \jshelter{} tries to mitigate.

\subsection{T1: User tracking}

In theory, laws like GDPR and ePrivacy Regulation give each person control over
their personal data and devices. However, there is a significant lack of
control over personal data on the
web~\cite{inria_cookie_banners,ico_adtech,WebTrackingPolicyTechnology,brave_complain,ryan_report}
in practice.
The advertisement technologies are under considerable scrutiny in Europe
\cite{belgianDPATCF,norwegianDPA}, but tracking scripts are omnipresent on the web.
Users risk complete disclosure of their browsing history.

Historically, trackers stored user identifiers in third party cookies. However,
browser vendors limit third party cookies. Hence trackers move to alternative
ways of identifying users. Browser and device fingerprinting is a stateless tracking method that
tries to find features that make (almost) every browser uniquely
identifiable~\cite{browser_fingerprinting,beauty_beast,crossbrowser_fingerprinting,fingerprinting_survey}.
For example, the content of HTTP headers, including user agent string, screen
size, language, time zone, and system fonts, together with hardware-dependent
characteristics such as canvas image
rendering~\cite{canvas_fingerprint,crossbrowser_fingerprinting}, audio
processing~\cite{million_web_site_fingerprint}, installed
fonts~\cite{fingerprinting_fonts}, installed browser
extensions~\cite{jstemplates_property_traversal,xhound,extend_not_extend,latex_gloves}, the
sites that the user is currently logged in~\cite{extend_not_extend}, clock
skew~\cite{Kohno,jucs15} and other
techniques~\cite{fingerprinting_survey}. The goal of the fingerprinter is to
create a stable identifier of a user so that the user is identifiable on
different sites.
Device fingerprint is the same in every browser on the same device, while
browser fingerprint differs for different browsers running on the same device.
Recent studies have shown that user tracking is
becoming more prevalent and complex~\cite{internet_jones}.
Note that the leaking information may uncover vulnerabilities of the
fingerprinted systems, and a fingerprinting database can be a valuable source of
information for an adversary wanting to misuse the data.

\begin{figure}[h!]
  \centering
    \includegraphics[width=0.48\textwidth]{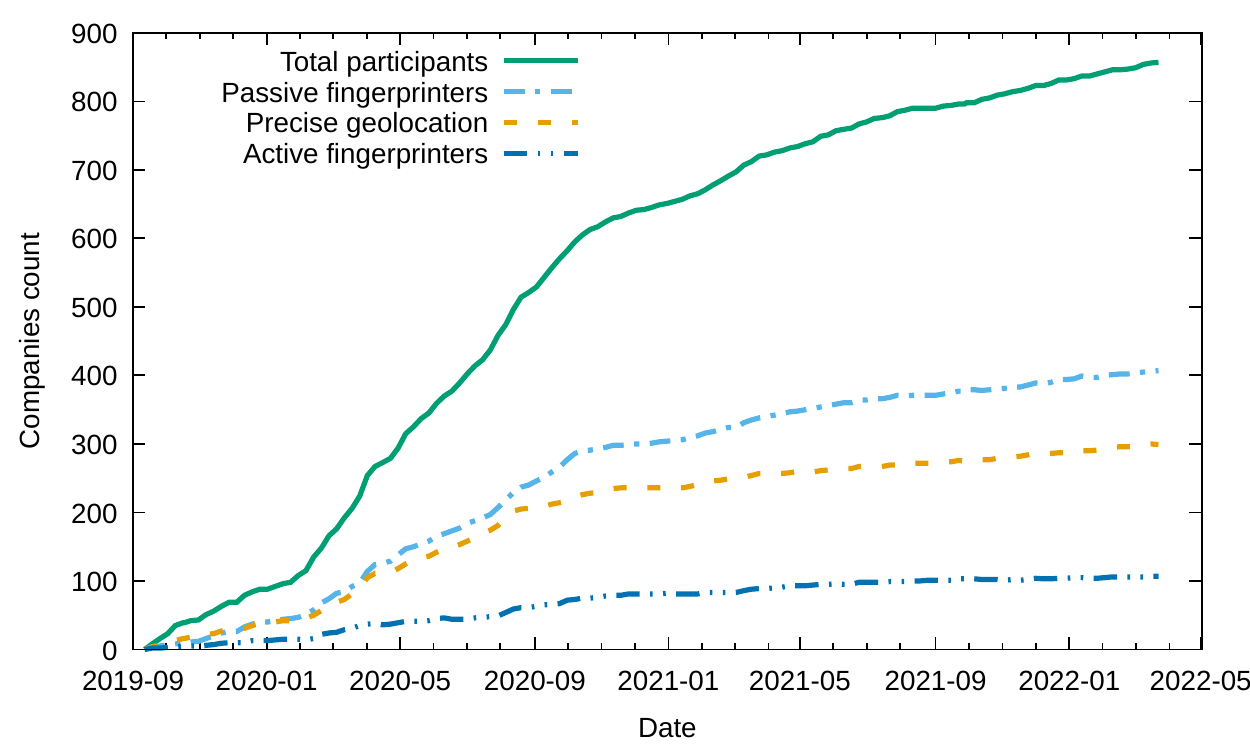}
  \caption{TCF participants reporting fingerprinting
  activities and
  precise geolocation processing.}
 \label{fig:tcffingerprinting}
\end{figure}

Fingerprinting is considered passive when it
contains information from HTTP headers or network traffic that is exchanged
regardless of whether the fingerprinting is in place.
On the other hand, active fingerprinting runs JavaScript code to retrieve data from
browser APIs.
Figure~\ref{fig:tcffingerprinting} depicts publicly
available data by IAB Europe Transparency and Consent Framework (TCF)\footnote{See
\url{https://vendor-list.consensu.org/v2/archives/vendor-list-vNUM.json} where
NUM is the number of the week since the start of the framework. See
\url{https://www.fit.vutbr.cz/~polcak/tcf/tcf2.html} for more data from the
framework.}. TCF allows companies to self-report active and passive
fingerprinting. More than
400 companies passively fingerprint users and more than 100 companies actively
use JavaScript APIs to create a unique fingerprint.
One of the goals of \jshelter{} is to prevent active fingerprinting.

Several studies monitored the deployed fingerprinting techniques on the
Internet~\cite{web_never_forgets,fpdetective,cookieless_monster,million_web_site_fingerprint,hiding_crowd,fingerprint_fingerprinters}.
Trackers exploit evercookies, shared cookies, font enumeration, canvas, web
audio, WebRTC, and many other APIs
to identify browsers and their users.
However, current countermeasures
are insufficient for a dedicated
fingerprinter willing to reveal inconsistencies in API readings~\cite{fpscanner,jstemplates_property_traversal}.

Current research distinguishes targeted and non-targeted
fingerprinting~\cite{fingerprinting_survey}. Non-targeted fingerprinting focuses
on observing visiting browsers or device fingerprints and trying to link their
identity to a previous visitor. Targeted fingerprinting tries to detect a
tailored fingerprint of an individual, for example, for law enforcement
investigations~\cite{jstemplates_property_traversal}.

Browser fingerprinting can also be used for benign use cases like multifactor
authentication --- if a website detects that a user connects from the same
device as previously seen, it is not necessary to perform additional
authentication steps. A website can recommend installing critical security
updates based on system properties, like the browser version. Some
websites collect browser fingerprints to distinguish between human users and bots to
prevent fraud.


Another branch of previous literature focused on processing behavioural biometric features derived from input
user interaction (keyboard, mouse, and touch events). For example, it is
possible to identify
users uniquely~\cite{MouseDynamicsAuthentication,MouseMovementUserVerification}, derive
handedness~\cite{HandednessKeystroke}, or age and
gender~\cite{InputEventsAgeGender}.

\subsection{T2: Very rich browser APIs}

Modern websites offer the capabilities of native applications. Browsers
support video calls, audio and video editing, maps and navigation,
augmented and virtual reality. One can control games with gamepads or check the
battery status. The web page may change its appearance according to the ambient
light. Nevertheless, most web pages do not need these advanced
APIs~\cite{webapi-vibrate}. Service Worker API allows Man-in-the-Middle
adversaries inject long-lasting trackers~\cite{nelsecrypt23}. Some APIs like Geolocation or microphone and
camera access need explicit approval by users. Others like gamepads, virtual reality,
battery, or sensors are available for
all visited pages\footnote{Sensor APIs are currently implemented, or partially implemented, in
Chromium-based browsers like Chrome, Edge, and Opera. For Android devices, the
support exists in Chrome for Android, Opera for Android, and various
Chromium-based browsers like Samsung Mobile or Kiwi Browser. The concrete
support for individual classes depends on the browser type and version. Some
features are considered experimental and only work when browser flags like
\texttt{\#enable-experimental-web-platform-features} or
\texttt{\#enable-generic-sensor-extra-classes} are enabled. Sensor APIs are
enabled by default in Chrome on Android.}

Iqbal et al.~\cite{fingerprint_fingerprinters} detected misusing the APIs by
many fingerprinting scripts.
Both Generic Sensor W3C Candidate Recommendation
Draft\footnote{\url{https://www.w3.org/TR/2021/CRD-generic-sensor-20210729/\#main-privacy-security-threats}}
and literature mention several risks stemming from sensor reading like location tracking~\cite{accomplice}, eavesdropping,
keystroke monitoring, device fingerprinting~\cite{accelerometer_fingerprinting},
and user identification~\cite{walking_patterns}.
Fig.~\ref{fig:tcffingerprinting} reports TCF data on companies using
precise geolocation data (precision better than 500 meters).

\subsection{T3: Local network scanning}

Devices browsing the web are typically
connected behind NAT, which does not allow external hosts to open connections to
devices in the local network (e.g. printers). Although the same-origin policy does
not allow a web page to access arbitrary resources, there are
side channels that might provide enough information about an existence of a
resource, including resources in the local
network~\cite{forcepoint}. A web page can try to exploit the web browser as
a proxy between the remote website and resources in the local network.
Bergbom~\cite{forcepoint} demonstrated that it is possible to execute arbitrary
commands on a local machine under certain circumstances (in this case, it was an
insecure Jenkins configuration).

\subsection{T4: Microarchitectural attacks}

Previous research also focused on side-channel attacks that can reveal what the user
has recently done with the device. For example, content-based
page deduplication performed by an operating system or a virtual machine hypervisor
can reveal if specific images or websites are currently
opened~\cite{gruss_pagededuplication} on the same device (hardware), possibly on
another virtual machine. The reply time for a specific request depends on the
cached content, so the reply time reveals if the content was recently
visited~\cite{timing_cache}. Moreover, even uncached content leaks information
on the server state~\cite{timing_http}. Bortz and Boneh~\cite{timing_http}
studied server reply times influenced by different code paths taken by the
server and were able to reveal private information. The
\verb|requestAnimationFrame| API can be used to time browser rendering
operations and reveal information on browser history and read pixels from
cross-origin iframes~\cite{timing_requestAnimationFrame}.

Operating systems
isolate processes from each other and the kernel. However, deficiencies in
hardware can provide possibilities to circumvent the isolation. Gruss et
al.~\cite{gruss_rowhammer_js} exploited JavaScript to modify memory cells
belonging to different processes (the attack is called Rowhammer). Hence, they
gained unrestricted access to systems of website visitors. They exploited
operating systems optimisations and high-precision
timings~\cite{gruss_rowhammer_js}. Later, Gruss et al.~\cite{gruss_rowhammer}
showed that industry countermeasures against Rowhammer attacks are ineffective.
Spectre attacks can be executed from JavaScript and leak data in the memory of other processes
running on the same system~\cite{Spectre}.

Some websites provide different content based on age, gender and location.
Van Goethem et al.~\cite{clock_still_ticking} employed timing attacks to reveal
data about users by measuring the size of the reply for resources
with different contents for different users.

Smith et al.~\cite{history_revisited} exploited
visited link pseudoclass and timed link redrawing based on the target URL to
determine browser history.

\section{Countermeasures}
\label{sec:countermeasures}

Many popular security and privacy-enhancing approaches already exist.
Let us focus on existing tools addressing the threats raised in Sect.~\ref{sec:threats}.

\subsection{Browser extensions}

Adblockers and other tracker blockers employ
lists of URLs or parts of URLs
that are considered harmful to user privacy or security. The advantage for the user is
that many tools focus on blocking (for example, uBlock Origin, EFF
Privacy Badger, Ghostery) and also blocklists that are usually
compatible with several blockers. Browsers like Firefox~\cite{FirefoxTP} and Brave include
tracking prevention by default.
The downside is that it is easy to evade blockers~\cite{block_me_if_you_can}.
The malicious web server needs to
change the URL of the script.
Hence, blocklists are very useful as a first-line defence and improve web
performance~\cite{FirefoxTP}. \jshelter{} users should install a tracker
blocker. However, blockers are not enough as the niche cases
evade the blockers~\cite{block_me_if_you_can}.

Webextensions like NoScript Security Suite\footnote{\url{https://noscript.net/}}
and uMatrix Origin\footnote{\url{https://github.com/gorhill/uMatrix}} allow users to
block JavaScript or other content either completely or per domain. Hence, they
can address all threats raised in Sect.~\ref{sec:threats}. However,
many pages depend on JavaScript. Users must select what content to trust.
HTTP Archive
reports
that an average page includes 22 external requests (21 requests for mobile devices)\footnote{\url{https://httparchive.org/reports/page-weight?start=earliest&end=latest&view=list\#reqJs}}
So a user trying to
determine what code to run
requires excellent knowledge about the external sites. Moreover, a malicious code may be only a
part of a resource; the rest of the resource can be necessary for correct page
functionality. So we believe that webextensions like NoScript Security Suite and uMatrix
Origin are good but do not protect the user from accidentally allowing
malicious code.

JavaScript Zero~\cite{js0} (also known as Chrome Zero\footnote{\url{https://github.com/IAIK/ChromeZero}}) expects
that a user lets the browser run the vulnerable code and focuses on mitigating
T4. Even
most skilful users can run malicious code if the script URL evades blocklists and other parts of the
code are needed for the page to display correctly.
However, the practical implementation supports only
Chromium-based browsers, is not maintained since 2017, and Shusterman et
al.~\cite{PrimeProbe1JS0} have
shown that the web page can obtain access to the original API calls.

Web API Manager \cite{webapi-vibrate} classifies JavaScript APIs into 81
standards\footnote{\url{https://github.com/snyderp/web-api-manager/tree/master/sources/standards}}.
A Web API Manager webextension user can disable all functionality defined
by any of the standards. The authors prepared three configurations
with standards blocked depending on their benefits and
costs~\cite{webapi-vibrate}. Web API Manager is most effective against T2 and
not targeted T1, but it can help mitigate other threats. Unfortunately, Web API Manager does not allow a user
to disable only a part of the standard, e.g. it is not possible to enable Canvas
API for drawing but disable reading. Note that canvas
fingerprinting is based on the reading~\cite{canvas_fingerprint}. Additionally, the webextension is no
longer maintained\footnote{See the message on the GitHub page
\url{https://github.com/snyderp/web-api-manager/blob/master/README.md}}, it is
not compatible with Firefox Multi-Account Containers\footnote{See
\url{https://github.com/snyderp/web-api-manager/issues/53} for more details},
and it suffers from the Firefox bug related to Content Security Policy (CSP)~\cite{ff-bug-csp}. A Web API
Manager user with a tailored configuration can potentially be uniquely
identified with the JavaScript enumerating code developed by Schwarz et
al.~\cite{jstemplates_property_traversal}.

As discussed below, we also studied two fingerprinting detection extensions,
namely A Fingerprinting Monitor For Chrome 
(FPMON)~\cite{fpmon} and Don’t FingerPrint Me
(DFPM)\footnote{\url{https://github.com/freethenation/DFPM}}. Both extensions
are not maintained anymore and are not available in webextension stores.

\subsection{Privacy-focused browsers}

Tor is a network of onion routers that allow relaying TCP connections so that
the server does not learn the IP address of a client but an IP address of their Tor
exit node. Tor Browser is a Firefox fork that tries to make every instance as
uniform as possible. For example, every user should browse with the same window
size. However, a fingerprinter can still learn some information like the underlying
operating system~\cite{fingerprinting_survey}. Tor Browser also disables several
APIs like WebGL. Consequently, Tor Browser is a very good solution to tackle
threats T1, T2, T3, and T4.

Nevertheless, Tor Browser users should not
resize the window and install
additional webextensions. These requirements downgrade comfort, and users might be unwilling to
abandon favourite webextensions or be tempted to resize the window for more comfort. As the
communication is relayed multiple times by relays spread worldwide,
latency increases, and throughput is limited.
Moreover, malicious actors often misuse Tor.
The list of Tor exit node IP addresses is public.
Some services block Tor
traffic, either to prevent frequent attacks or as a temporary measure to block
an attack.

Brave browser is a Chromium fork that focuses on privacy. For example, it has a
built-in blocker and anti-fingerprinting solution. Using Brave is a good option
to tackle T1, T3, and T4. A disadvantage is the long build time. Often, it
is not available in GNU/Linux distribution repositories.

Mozilla is working on integrating fingerprinting resisting
techniques from
Tor Browser\footnote{\url{https://bugzilla.mozilla.org/show_bug.cgi?id=1329996}}
to Firefox (Firefox Fingerprinting Protection, also known as resist
fingerprinting).
However, the work is not done, and it is a possible related research question
if the hiding in the herd strategy makes sense before it is adopted for all
users. Moreover, inconsistencies arise. For example, Tor Browser does not implement
WebGL. As Firefox adopts fingerprinting protections from Tor Browser, Firefox modifies readings
from a 2D canvas and does not modify a WebGL canvas. That creates a false sense of
protection.

\subsection{Current browser fingerprinting countermeasures}
\label{sec:3antifp}

Modifying the content of fingerprints is a valid choice to resist a
fingerprinting attempt. However, each modification may create an inconsistency
that may improve the fingerprintability of the
browser~\cite{fingerprinting_survey}. Currently, three anti-fingerprinting
approaches exist.

(1) \emph{Create homogeneous fingerprints}. If the commonly used fingerprinting
APIs returned the same values in every browser, a fingerprinter would not be
able to construct a fingerprint and tell the users behind the browsers apart. The
leading representative of this approach is Tor Browser. Unfortunately,
homogeneous fingerprints have an inherent downside of following specific rules
to be effective. Most importantly, the effectiveness of the approach depends on
the broad coverage of the blocked APIs and the size of the population employing
the countermeasures. All browsers with the same fingerprint form an anonymity
set \cite{anon_terminology}. An observer cannot distinguish between browsers in
the anonymity set. With every missed fingerprintable attribute, the anonymity
set breaks into smaller sets. For example, Tor Browser strongly recommends
using a specific window size. Suppose a user changes the window size to a value
different from all
other Tor Browser users. In that case, a fingerprinter can identify the user solely
by this attribute. Moreover, Tor Browser hides the IP address of the user. A
webextension cannot hide or mask the IP address.

(2) \emph{Change the fingerprints on different domains to disable cross-domain
linkage}. Brave browser modifies the results of APIs commonly used for
fingerprinting. Its goal is to create a unique fingerprint for each
domain and session. As the output of APIs commonly used for fingerprinting
changes for every visited domain, it cannot be used for cross-domain linking of
the same browser.

(3) \emph{Detect and block fingerprint attempts}. A protection tool can monitor
access to properties commonly misused for fingerprinting and block access to
additional properties or limit the page ability to upload the fingerprint.  To
reliably prevent sharing the fingerprints with trackers, any network traffic to
the tracking server has to be blocked, and the web page cannot have an
opportunity to store the fingerprint for retrieval after page reload. Such
measures can be effective against fingerprinting. Nevertheless, they also impose
severe restrictions on web applications, limit overall usability, and break
page behaviour. Fingerprinting detection can also be imprecise. In practice, it
takes time to detect that a fingerprint is indeed being computed. As a page can
immediately send the values being read for fingerprinting to the server, the
server can learn a partial fingerprint before the fingerprinting is detected and blocked.

\section{\jshelter{} design decisions}
\label{sec:methodology}

As the current state-of-the-art covered in Sect.~\ref{sec:countermeasures}
suggests, there is no  perfect and straightforward solution for the threats raised in
Sect.~\ref{sec:threats}. This section covers the design decisions of \jshelter{}
and the countermeasures we decided to implement.

\jshelter{} goals are
as follows:

\begin{enumerate}

  \item Create a webextension because webextensions work across multiple
    browsers and consequently can be easily installed into any browser that
    supports webextensions, including Firefox and all browsers based on Chromium.

  \item Do not create a perfect solution, instead focus on what other
    webextensions lack: a consistent approach to the threat T1 and protection
    from T2, T3, and T4.

  \item Let people with different knowledge depths use the extension.

\end{enumerate}

Chrome Zero~\cite{js0} and Web API Manager~\cite{webapi-vibrate} were the
inspiration for \jshelter{}. Chrome Zero applies protection based
on closures and Proxy objects and focuses on microarchitectural attacks. Web API
Manager provides a way to disable browser APIs selectively. Both Chrome Zero and
Web API Managers are no longer maintained.

Currently, \jshelter{} offers three types of protections. (1)~JavaScript Shield
(JSS)
modifies or disables JavaScript APIs. It aims at threats T1, T2 and
time-measurement-related
protection for T4.
(2)~Fingerprint Detector (FPD) provides heuristic analysis of fingerprinting behaviour
and tackles T1, (3)~Network Boundary Shield (NBS) monitors the source and destination of
each web request and detects attempts to misuse the browser as a proxy to the local
network (T3).

\subsection{Fingerprint Detector}

Fingerprint Detector (FPD) monitors APIs that are commonly used by
fingerprinters and applies a heuristic approach to detect fingerprinting
behaviour in real-time (see threat T1). 
When a fingerprinting attempt is detected, FPD notifies the user. The user can configure \jshelter{} to reactively block subsequent asynchronous HTTP requests initiated by the 
fingerprinting page and clear the storage facilities where the page could have stored a (partial) fingerprint. However, this behaviour may break the page. The goal of the aggressive 
mode is to prevent the page from uploading the full fingerprint to a server.
However, the fingerprinter can gradually upload detected values, and a partial fingerprint can leak 
from the browser.

We chose the heuristic approach as many prior studies \cite{fingerprinting_survey,web_never_forgets,million_web_site_fingerprint,fingerprint_fingerprinters} 
proved it to be a viable approach with a very low false-positive rate.
\jshelter{} heuristics count calls of JavaScript API endpoints, which are relevant for 
fingerprinting detection, performed by a web page.
FPD is not based on code analyses, so it overcomes any obfuscation of 
fingerprinting scripts. FPD builds upon recent knowledge in the
field:

(1) Iqbal et al.~\cite{fingerprint_fingerprinters} measured the relative prevalence of API keywords in fingerprinting scripts and created a list of APIs using the ratio metric. 
FPD observes wrappable endpoints that are frequently misused (at least 10
occurrences counted by Iqbal et al.).

(2) Additionally, FPD includes heuristics proposed by Englehardt and Narayanan~\cite{million_web_site_fingerprint} to detect additional fingerprinting techniques. These techniques often require more 
steps to produce a fingerprint (e.g. canvas fingerprinting). FPD heuristics use groups to define these steps regardless of the order. 

(3) We looked through the source code of fingerprinting tools like FingerprintJS\footnote{\url{https://github.com/fingerprintjs}}, Am I Unique\footnote{\url{https://amiunique.org/}} and 
Cover Your Tracks\footnote{\url{https://coveryourtracks.eff.org/}}. Furthermore, we analysed 
FPMON~\cite{fpmon} and DFPM\footnote{\url{https://github.com/freethenation/DFPM}}. Finally, we pick all 
the relevant endpoints and group them by their semantic properties. We further
adjusted the weights during testing.

\begin{figure}[h!]
  \centering
    \includegraphics[width=0.45\textwidth]{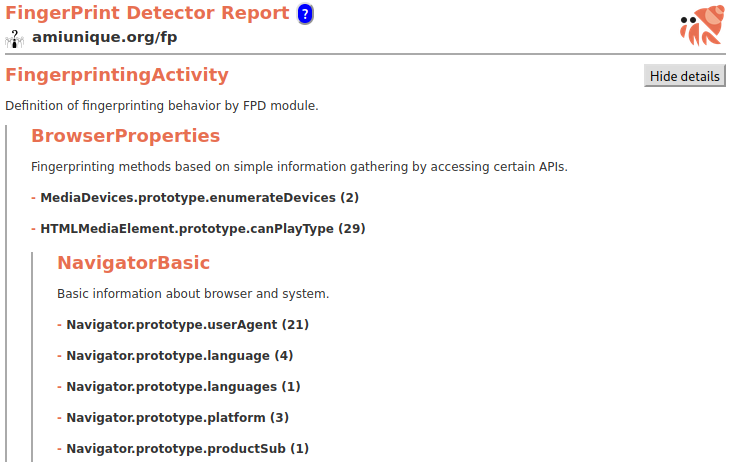}
  \caption{An excerpt from an FPD report on \url{AmIUnique.org}. The user can clearly see what APIs the
  visited page called.}
 \label{fig:fpd}
\end{figure}

FPD provides a report that summarises FPD findings on the visited web page, see Fig.~\ref{fig:fpd}. The report aims to educate users about fingerprinting and clarifies why FPD 
notified the user and optionally blocked the page. Additionally, the report can
be generated from passive observation of a web page (no API blocking). We expect
that other researchers will use passive FPD to study fingerprinting in more detail.

We expect that the APIs for fingerprinting will change in time so we designed the heuristics to be as flexible as possible. We expect to run periodic web crawls 
based on the tools initially developed by Snyder et al.~\cite{Snyder_features} and apply machine learning to FPD.

\subsection{JavaScript Shield}
\label{sec:jss}

JSS focuses on spoofing timestamps (threat T1 and T4), fingerprint modifications
(threat T1) and limiting APIs available to visited pages (threat T2).
JSS offers three predefined profiles that we expect users should use.

Profile P1 focuses on making the browser appear differently to distinct
fingerprinting origins by slightly modifying the results of API calls
differently on different domains so that the cross-site fingerprint is not
stable~\cite{PriVaricator,FPRandom}. The focus is on applying security countermeasures that are likely not to
break web pages. However, as some modifications break some pages. If FPD does
not detect fingerprinting attempt, \jshelter{} allows users to apply another
profile (P2) that does not apply fingerprinting protections but incorporates
other security protections.

Profile P3 focuses on limiting the information provided by the browser by
returning fake values from the protected APIs. Some are blocked completely, some
provide meaningful but rare values, and others return meaningless values.

P2 and P3
make the user fingerprintable because the results of API calls are generally
modified in the same way on all websites and in each session.

\jshelter{} currently modifies 113 APIs, which include APIs considered by
previous works of Schwarz et al.~\cite{js0},
Iqbal et al.~\cite{fingerprint_fingerprinters}, Snyder et
al.~\cite{webapi-vibrate} and APIs that Apple declined to implement. For each API, we decide
its relevance on an individual basis. Usually, we do not modify APIs
already explicitly permitted by the user. However, the analysis might provide an example
where the user still wants to limit the precision of the API. For example,
Geolocation API allows the page to learn a very precise location while the user might be
interested in services in the city. Hence, \jshelter{} allows fine-tuning the
precision of the Geolocation (and other APIs).

Additionally, the slightest mismatch between the results of two APIs can make the user more visible to
fingerprinters~\cite{fingerprinting_survey,everyone_different,jstemplates_property_traversal}.
Hence, we consider each protection that we decide to implement in \jshelter{}
from the point of fingerprintability, the threat
of leaking information about the browser or user and other threats presented in
Sect.~\ref{sec:threats}. \jshelter{}
tries to mimic a stationary device with consistent and plausible readings.

\subsubsection{Farbling-like prevention of browser fingerprinting}
\label{sec:farbling}

JSS applies the same or very similar anti-fingerprinting modifications as Farbling implemented in
Brave\footnote{See \url{https://github.com/brave/brave-browser/issues/8787} and
\url{https://github.com/brave/brave-browser/issues/11770}}.
Farbling is, in turn, based on
PriVaricator~\cite{PriVaricator} and FPRandom~\cite{FPRandom}. JSS modifies the values readable
by page scripts with small lies that differ per origin.
These little lies result in different
websites calculating different fingerprints. Moreover, a previously visited
website calculates a different fingerprint in a new browsing session.
Consequently, cross-site tracking is more complicated.

Datta et al.~\cite{evaluating_afpets} evaluated several anti-fingerprinting
approaches. Nevertheless, \jshelter{} was not evaluated by Datta et al. as the
project did not exist during the time they performed their study. \jshelter{}
would not apply many modifications to properties studied by Datta et al. because
Datta et al. focused on properties that allow determining browser and operating
system versions. We do not see a way to consistently spoof operation system,
browser, and the version from a web extension. Consequently, \jshelter{} does
not try to partially spoof such information as that would make the browser more
fingerprintable. Indeed, Datta et al. found all webextensions not masking more
than 50\,\% of the studied attributes. Additionally, Datta et
al.~\cite{evaluating_afpets} prefer the approach of Torbrowser.
Their expectation is that one anti-fingerprinting approach \emph{"could become
nearly universal in the future"}. We develop \jshelter{} for the current web
where it is used only rarely. Furthermore, Datta et al.~\cite{evaluating_afpets}
do not provide a sound comparison of homogeneous fingerprints and little lies as
they \emph{"assume any modifications of an attribute renders that attribute
useless to a tracker"}. We expect that a fingerprinter can use the spoofed
values in the fingerprint so our approach is to confuse the fingerprinter with
little lies.

\subsubsection{Interaction between JavaScript Shield and Fingerprint Detector}

Both JSS and FPD aim to prevent fingerprinting. Both are necessary for
\jshelter{}.

\begin{itemize}

  \item The blocking mode of FPD breaks pages. Users are typically tempted
    to access the content even when they know they are being fingerprinted.
    Consequently, they turn FPD off for such pages. JSS ensures that these users
    are not linkable across origins and sessions.

  \item JSS profiles P2 and P3 likely
    provide the same fingerprint for all domains; hence, we strongly advise
    users of this profile to activate FPD.

  \item We expect most users to stick with the default profile
    creating little lies. Future research should validate the current
    approach. For example, \jshelter{} and Brave create indistinguishable
    changes to canvas readings. These are sufficient for a fingerprinter that
    creates a hash of the readings. Nevertheless, an advanced fingerprinter might, for
    example, read the colours of specific pixels to determine a presence of a font
    (different fonts produce a different pixel-wise-long output of the same text).
    As both Brave and \jshelter{} modify only the least significant bit of each
    colour, the fingerprinter can ignore this bit and get the information on
    installed fonts. Hence, we believe that FPD is beneficial as it offers
    additional protections.

\end{itemize}

\subsubsection{Sensors}

\jshelter{} tries to simulate a stationary device and consequently completely
spoofs the readings of AmbientLight, AbsoluteOrientation,
RelativeOrientation, Accelerometer, LinearAcceleration, Gravity, Gyroscope, and
Magnetometer sensors.
\jshelter{} also spoofs Geolocation API that can be either completely blocked or
return a modified location derived from the reading from the original API.

Instead of using the original data, \jshelter{} returns artificially generated
values that look like actual sensor readings. Hence the spoofed readings
fluctuate around a value that is unique per origin and session. The readings are
performed consistently in the same origin tabs, so the same sensor produces the
same value in each tab.

We observed sensor readings from several devices to learn the fluctuations of
stationary devices in different environments. Most of the sensors have small
deviations. However, magnetometer readings have big fluctuations.
\jshelter{} simulates Magnetometer fluctuations by using a series of sines for each axis.
Each sine has a unique amplitude, phase shift, and period.
The number of sines per axis is chosen pseudorandomly.
\jshelter{} currently employs 20 to 30 sines for each axis.
Nevertheless, the optimal configuration is subject to future research.
More sines give less predictable results at the cost of increased computing
complexity.

The readings of the acceleration and orientation sensors are generated
consistently between each other from an initial device orientation that
\jshelter{} generates for each origin and session.

\subsubsection{User in Control}

The number of modified APIs is high. We expect that users will encounter pages
broken by \jshelter{} or that do not work as expected. For example, the
user might want to play games with a gamepad device on some pages or make a call
on others.

\begin{figure}[h!]
  \centering
    \includegraphics[width=0.45\textwidth]{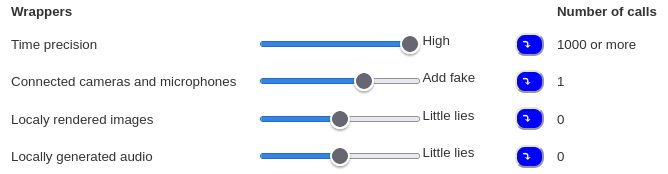}
  \caption{JSS reports back which APIs are being used by the page.}
 \label{fig:jss}
\end{figure}

JSS allows each user to fine-tune the protection for each origin.
Some users reported that they would prefer to avoid digging into the
configuration. Those can disable JSS for the domain with a simple ON/OFF popup
switch. More experienced users can react to information provided by FPD and turn
off JSS fingerprint protection when the visited site does not behave as a
fingerprinter. The most experienced users can fine-tune the behaviour per API group.
Figure~\ref{fig:jss} shows an example of a user accessing a page that allows
video calls. The user sees the groups with APIs that have been called by the
visited page at the top and can
quickly fix a broken page.

\subsection{Effective modifications of the JavaScript environment}

Monitoring and modifying the results of the
built-in JavaScript APIs and built-in object behaviour
is the \jshelter{} core functionality. \jshelter{} employs the
same mechanism proposed by Schwarz et al.~\cite{js0} in Chrome Zero. However,
Chrome Zero was a proof-of-concept with no modification in the last four years.
Shusterman et al.~\cite{PrimeProbe1JS0} identified several problems with Chrome Zero:

\begin{enumerate}

  \item Unprotected prototype chains (\emph{issue~1}): the original implementation is available
    through the prototype chain because Chrome Zero protects a wrong property.

  \item Delayed JavaScript environment initialisation (\emph{issue~2}): Current webextension APIs lack a
    reliable and straightforward way to inject scripts modifying the JavaScript environment before
    page scripts start running. \jshelter{} and Chrome Zero allow configurable protection
    that may differ per origin, so they need to load
    the configuration during each page load. Hence, a naïve implementation with
    asynchronous APIs may allow
    page scripts to access original, unprotected API calls. Note that once page
    scripts can access the original
    API implementation, they can store the unprotected version. There is no
    way (for a webextension) to reverse the leak.

  \item Missed context (\emph{issue~3}): Chrome Zero does not apply protection in iframes and
    worker threads.

\end{enumerate}

In addition, Firefox suffers from a
long-standing unfixed bug~\cite{ff-bug-csp}
that prevents up to 10\,\% of Firefox webextensions from working correctly on
pages whose Content Security Policy (CSP) forbids inline scripts
\cite{FirefoxCSPBugBP} (\emph{issue~4}).

To cope with issue 1, \jshelter{} needs to
identify the correct method to protect. For example, \verb|performance.now()|
yields precise timestamps. However, the method \verb|now| is not a method of the
\verb|performance| object, but it is rather available through prototype chain
from \verb|Performance.prototype|. We tackle this issue in two
steps. (1) Before we implement a protection, we analyse the prototype chain and pick
the correct object implementing the property or method to wrap.
Left alone, this approach is brittle: it can be broken by changes in the DOM APIs specification
or by browser implementation. Therefore, \jshelter{} applies an additional step.
(2)~The injection code checks at runtime whether the property (or method) is actually
implemented as an own property by the object defined in step 1 or if the
property is available through the prototype chain.
In the latter case, \jshelter{}
replaces the correct property by traversing the prototype chain and
overriding the step 1 choice.

To overcome issues 2--4, we needed to develop
a reliable cross-browser early script injection. As the same issues
affect several privacy and security webextensions, we refactored the code from \noscript{}
into \NSCL{}
and made it publicly available for reusing and contributing back.

\NSCL{} abstracts common functionality
shared among security and privacy webextensions. The goal is
to minimise the development and maintenance burden on webextension maintainers.
Shared common code paths between webextensions lead to more code review. As the
browser APIs often provide multiple ways to achieve the same goal, a shared
code between webextensions prevents feature mismatches. For example, an attacker
can access an API through the window object, an iframe, or a worker. A
webextension modifying the API needs to modify each possibility. By modifying
only some ways to access the API, the webextension not only gives an attacker the
possibility to learn original values offered by the API but also reveals that
the browser behaves strangely. Additionally, \NSCL{}
provides consistent
implementation across multiple browser engines. Hence, developers do not need to
study browser-dependent implementation details.

The \NSCL{} tackles issue 2 in its \verb|DocStartInjection|
module\footnote{\url{https://github.com/hackademix/nscl/20220330/main/service/DocStartInjection.js}}.
\NSCL{} allows to preprocess URL-dependent configuration inside a \verb|BeforeNavigate| event handler.
This event is fired every time the browser starts
loading a new page. As the event handler has access to the destination URL,
\jshelter{} can
build a configuration object in advance. Later, the \jshelter{} content script
can access the configuration during the document start event (before any page script can run).
However, this technique does not always succeed
due to race conditions.
As a safety net, when the content script finds no configuration object, it calls
\verb|SyncMessage| API
to retrieve the correct settings before it is interleaved
with concurrent scripts.

To address issue 3, \verb|manifest.json| (the configuration of the webextension)
registers code injection into all the newly created windows, including subframes.

Unfortunately, this alone cannot prevent dynamically created windows and frames from being
exploited by the originator page to retrieve unwrapped objects as
\verb|window.open()|,
\verb|contentWindow|, and \verb|contentDocument.window|
allow access to a new window object immediately after its creation
(synchronously) before any initialisation
(including the injection of webextension content script) occurs. To fix this
problem, \NSCL{}
\verb|patchWindow()| API modifies \verb|window.open()|,
\verb|contentWindow|, and \verb|contentDocument.window| to recursively wrap the newly created window just before it gets returned\footnote{\url{https://github.com/hackademix/nscl/blob/20220330/content/patchWindow.js\#L247}}.

A further possibility to access unwrapped APIs are subframe windows of all
kinds, also
immediately available at creation time by indexing their parent window as
an unwrappable pseudo array (e.g. \verb|window[0]| is a synonym of
\verb|window.frames[0]|). The \NSCL{} takes care of this problem by automatically patching
all not yet patched \verb|window[n]| objects every time the DOM structure is modified, potentially creating new windows.
This requires accounting for all methods and accessors by which the DOM can be changed in JavaScript and wrapping them\footnote{\url{https://github.com/hackademix/nscl/blob/20220330/content/patchWindow.js\#L311}}.

Regarding web workers, \jshelter{} disables them by default.
The \NSCL{} provides another option: wrapping
workers by injecting the wrappers in their own browser context via its
\verb|patchWorkers()| API. The implementation is very complex and still
experimental\footnote{\url{https://github.com/hackademix/nscl/blob/20220330/content/patchWorkers.js}
and \url{https://github.com/hackademix/nscl/blob/20220330/service/patchWorkers.js}}.
It needs more testing before it can be confidently deployed to a general audience.

Finally, \NSCL{} works around issue 4
by leveraging a Firefox-specific privileged API meant to safely share functions and objects between page scripts and
WebExtensions\footnote{\url{https://developer.mozilla.org/en-US/docs/Mozilla/Add-ons/WebExtensions/Sharing_objects_with_page_scripts}}.
On Chromium, where such API is not available, but injected scripts have no
special powers, and therefore, do not need those safety measures, \NSCL{} provides shims to expose a uniform interface for injected code and reduce the burden of cross-browser development.


\section{Evaluation}
\label{sec:results}

This section evaluates the different \jshelter{} parts.

\subsection{Fingerprinting inconsistencies}

Besides a few bugs that we intend to fix,
we are aware that a fingerprinter may observe some inconsistencies. For example,
\jshelter{} modifies each read canvas. Should the page scripts probe a
single-colour-filled
canvas, \jshelter{} would introduce small changes in some pixels. Hence, a
page script might learn that protection against canvas fingerprinting is in
place.

A naïve implementation available in earlier \jshelter{} versions modified all
canvases of the same size in the same way. Hence a fingerprinter could have created
two canvases, one for the fingerprinting and the other to learn what pixels are
modified and consequently revert the modifications. We removed the vulnerability
before anyone outside our team discovered the issue. Nevertheless, the little
lies modifications (see Sect.~\ref{sec:farbling}) have a
performance hit. For all APIs that allow obtaining hardware-rendered data like
the Canvas, WebGL, and WebAudio APIs, \jshelter{} needs to access all data in
two iterations, first to create a hash that controls the modifications in the
second iteration. Hence, the same content is deterministically modified the same
way, and different content is modified differently.

Consider \verb|AudioBuffer.prototype.getChannelData| allowing quick
access to pulse-code modulation audio buffer data without data copy.
A~fingerprinter might be interested in a couple of samples only. However, the
spoofing mechanism needs to access all data, so the method is much slower
(learning that the time of \verb|getChannelData| takes too long is
usable for fingerprinting).

We are not aware of any isolated side-effect that reveals
\jshelter{}. For example, page scripts can detect some similar webextensions by
calling \verb|Function.prototype.toString| for the modified APIs.
Should \verb|toString| return the wrapping code modifying the API rather than the
original value, it might reveal a unique text as other
webextensions modifying the same API call by the same technique will likely use a different
code. Nevertheless, we are aware and do not hide that users of \jshelter{} are
vulnerable to focused attacks. Our goal is to offer protections
indistinguishable from another privacy-improving tool for each modified API.
Nevertheless, a focused observer will very likely be always able to learn that a
user is using \jshelter{} if they aggregate the observable inconsistencies of
all APIs produced by \jshelter{}.

\subsection{Timing events}

\jshelter{} implements rounding and, by default, randomises the timestamps as
Chrome Zero does~\cite{js0}. In comparison, Firefox Fingerprinting Protection and
Tor Browser implement only rounding, which makes the technique visually easily
detectable. Compared with Chrome Zero, \jshelter{} modifies all APIs that
produce timestamps, including events (see threat T1), geolocation, gamepads,
virtual reality and sensors.

Computer clocks do not measure time accurately, but each has a built-in error.
Previous research~\cite{Kohno,jucs15,CS_experience} established that such errors
are unique to a device and observable on the network. Jireš~\cite{BPJires}
studied the influence of timestamp rounding (Tor Browser, Firefox Fingerprinting
Protection) and rounding and randomisation (\jshelter{}, Chrome Zero). He
computed clock skew from rounded timestamps but could not
remove the noise from rounded and randomised timestamps. However, this result
should be validated; long-lasting (at least tens of minutes)
measurements might remove the randomisation noise and reveal the clock skew.
Nevertheless, Polčák and Franková~\cite{jucs15} observed that timestamps
provided by JavaScript are affected by time synchronisations (such as NTP).
Hence, we advise combining \jshelter{} round and randomisation with
continuous-time synchronisation to hide built-in clock skew.

Biometrics cannot be forgotten or stolen~\cite{MouseDynamicsAuthentication}.
However, \jshelter{} forges timestamps from all JavaScript timestamp sources
consistently. As the biometric feature computation is based on time
tracking~\cite{MouseDynamicsAuthentication,MouseMovementUserVerification,HandednessKeystroke,InputEventsAgeGender},
forged timestamps result in fake biometrical data.

\subsection{Sensors}

\subsubsection{Sensor timestamp loophole}

\begin{figure*}[ht!]
  \centering
    \parbox{0.3\textwidth}{\centering
      \includegraphics[width=0.3\textwidth]{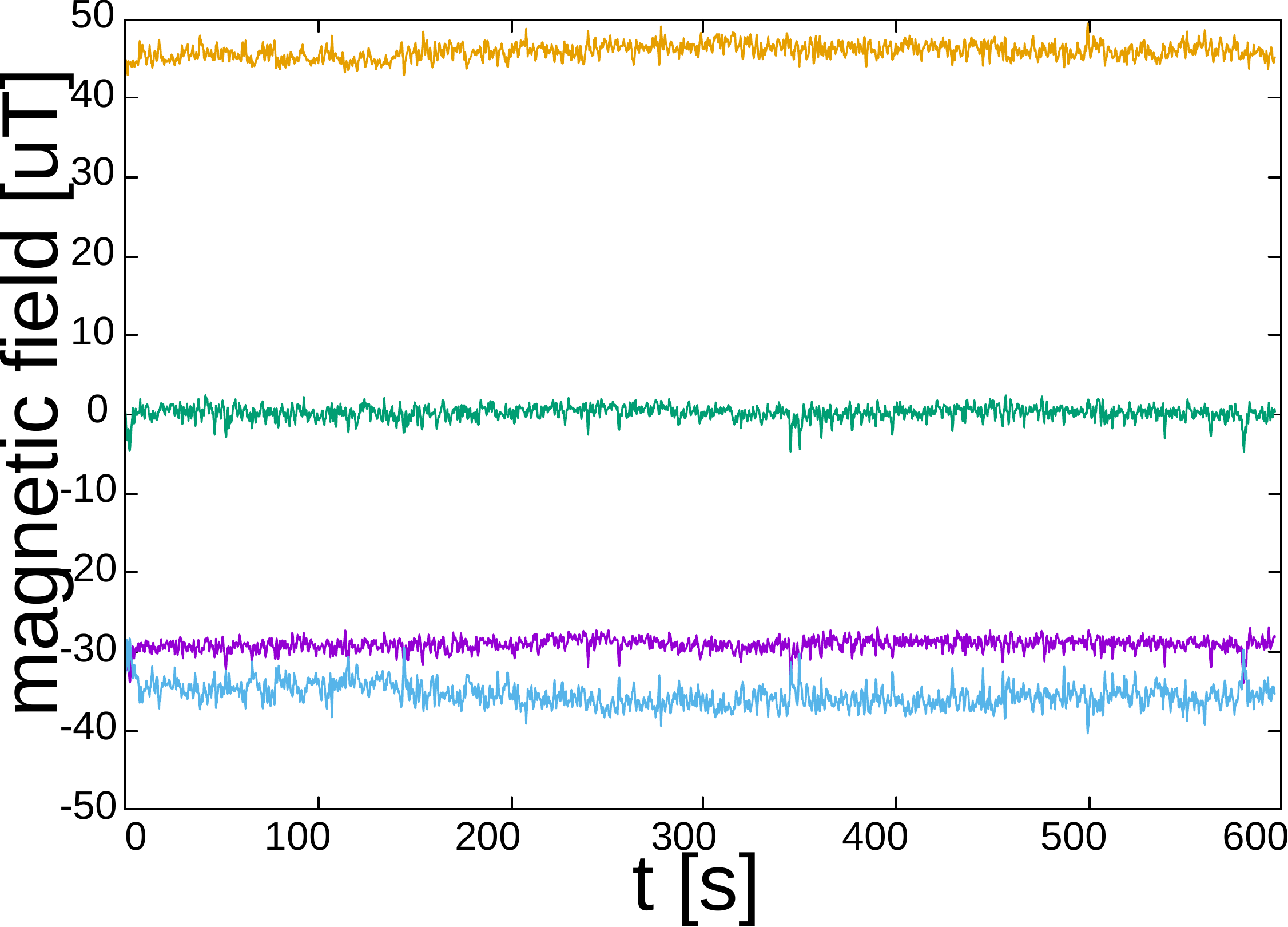}
      (a) Stationary device}
    \parbox{0.3\textwidth}{\centering
      \includegraphics[width=0.3\textwidth]{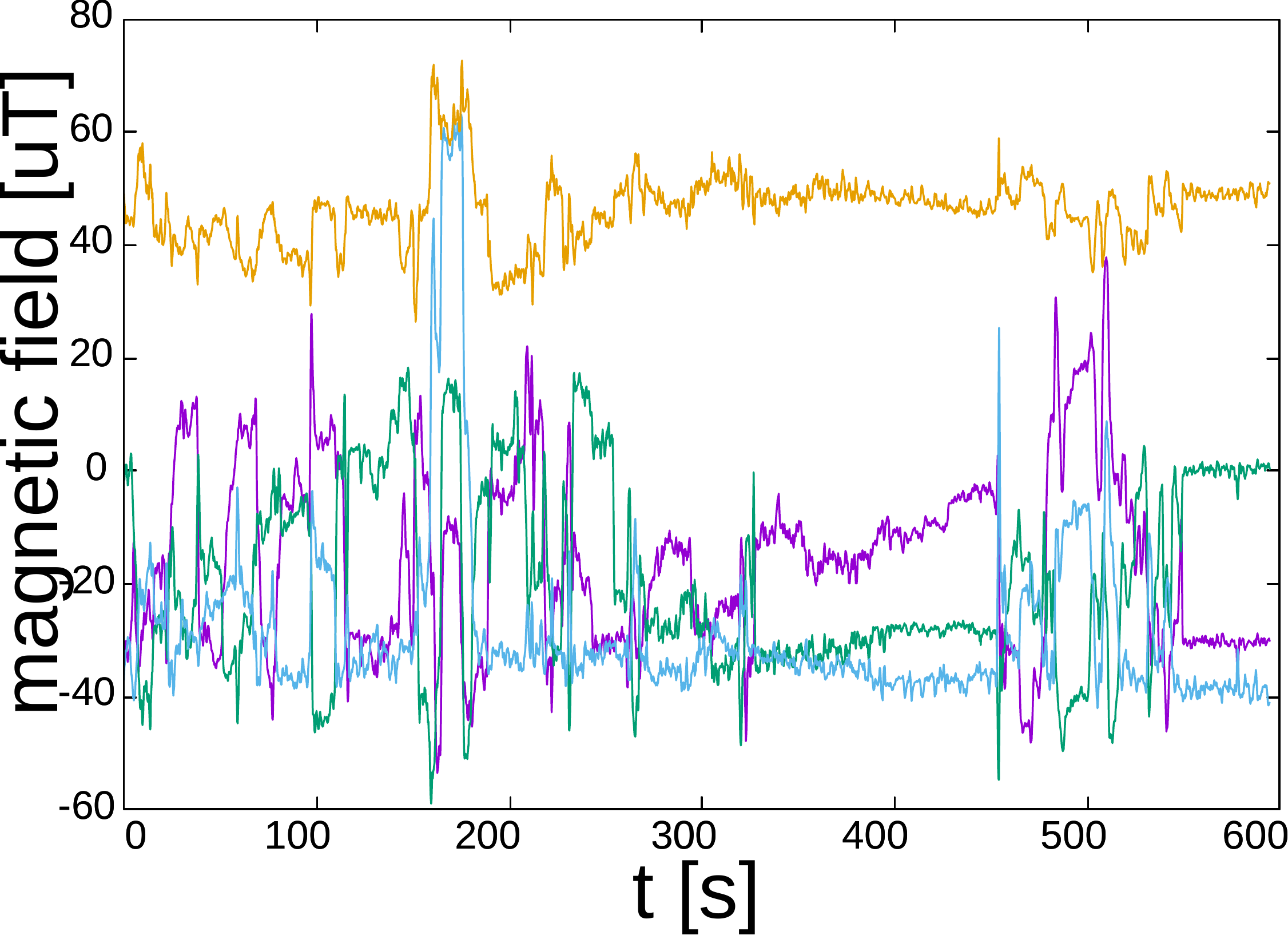}
      (b) Moving device}
    \parbox{0.3\textwidth}{\centering
      \includegraphics[width=0.3\textwidth]{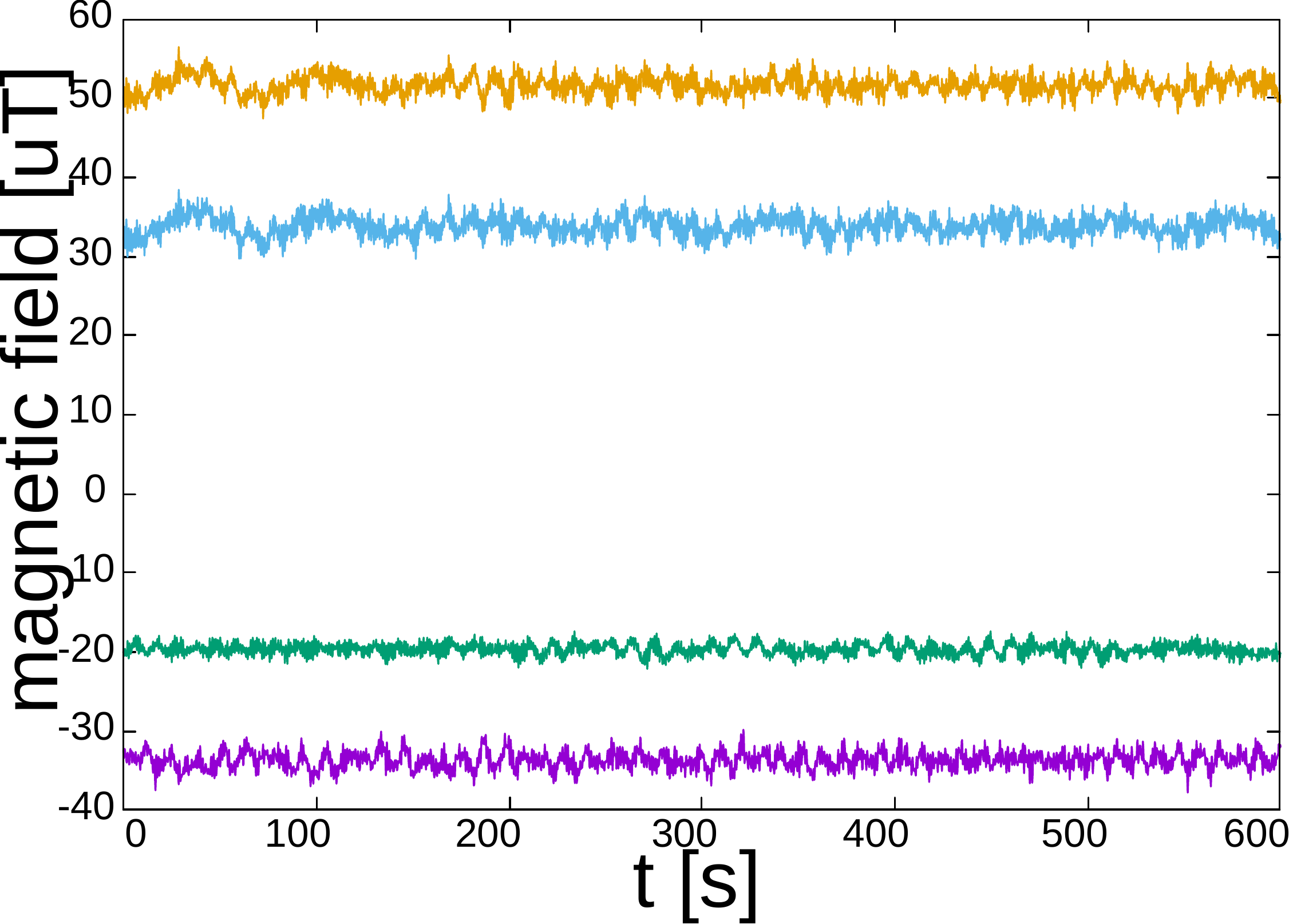}
      (c) Fake readings}
    \parbox{0.08\textwidth}{\centering
      \includegraphics[width=0.08\textwidth]{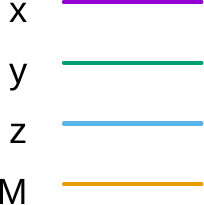}}
  \caption{Magnetometer readings.}
 \label{fig:magnetomer}
\end{figure*}

We discovered a loophole in the \verb`Sensor.timestamp`
attribute\footnote{Tested with Samsung Galaxy S21 Ultra; Android 11, kernel
5.4.6-215566388-abG99BXXU3AUE1, Build/RP1A.200720.012.G998BXXU3AUE1, Chrome
94.0.4606.71 and Kiwi (Chromium) 94.0.4606.56 and Xiaomi Redmi Note 5; Android 9,
kernel 4.4.156-perf+, Build/9 PKQ1.180901.001, Chrome 94.0.4606.71}. The value
describes when the last \verb`Sensor.onreading` event occurred in millisecond
precision. We observed that the reported time is the time since
the last boot of the device. Exposing such information is
dangerous as it allows fingerprinting the user easily as
devices boot at different times.

\jshelter{} protects the device by provisioning the time since the browser created
the page context (the same value as returned by \verb|performance.now()|. Such timestamps
uniquely identify the reading without leaking anything about a device. Future
work can determine if such behaviour appears in the wild. If all devices and
browsers incorporate the loophole, we should provide a random boot time.

\subsubsection{Fake magnetometer evaluation}

Figure~\ref{fig:magnetomer} shows readings from a real and fake magnetometer.
The left part (a) shows a stationary device. The magnetic field is not stable
due to small changes in Earth's magnetic field and other noise. The middle part of
the figure (b) shows a device that changed its position several times during the
measurement. We analysed traces of sensor readings collected in various
locations and environments. Fig.~\ref{fig:magnetomer} (c) shows readings
generated by \jshelter{} fake magnetometer. The values look like actual sensor
readings. Nevertheless, the generator uses a series of constants whose optimal
values should be the subject of future research and improvements.

\subsection{Fingerprint Detector effectivity}

The FPD heuristics were designed to keep the number of false positives as low as
possible. As FPD can optionally block all subsequent requests by a
fingerprinting page and \jshelter{} provides complementary protections, FPD
blocks only indisputable fingerprinting attempts. We conducted real-world
testing of FPD and refined its detection heuristics accordingly.

regarding testing methodology, we manually visited homepages and login pages of the
top 100 websites from the Tranco
list\footnote{\url{https://tranco-list.eu/list/23W9/1000000}}.
We randomly replaced inaccessible
websites by websites from the top 200 list. Before
visiting a website, we wiped browser caches and storage to remove
previously-stored identifiers. Hence, the visited pages may have deployed fingerprinting scripts more aggressively to identify the user and reinstall the identifier.

To boost the probability of fingerprinting even more, we switched off all
protection mechanisms offered by the browser. However, we blocked third-party
cookies because our previous experience suggests that the missing possibility to
store a permanent identifier tempts trackers to start fingerprinting. We
repeated the visits with both Google Chrome and Mozilla Firefox.

We used FPMON~\cite{fpmon}, DFPM, and \jshelter{} to find the ground truth. For each visited
page, we computed its fingerprinting score. FPMON reports fingerprinting pages
with colour. We assigned yellow colour 1 point and red colour 3 points. DFPM
reports danger warnings. If DFPM reports one danger warning, we assign 1 point
to the page. For a higher number of danger warnings, we assign 3 points to the
page. Therefore, each page gets a fingerprinting score from 0 to 6. We consider
each page with the score of 6 or 4 to engage in fingerprinting. Additionally, we
inspected pages with the score lower than 4 flagged by FPD. We detected five
additional fingerprinting pages after manual inspection.

Table \ref{tab:fpdstudy} shows the accuracy and the sum of true positives and
true negatives of the tested tools. In total, we tested
98 home pages and 81 login pages; 2 home pages are actually login pages, we
removed duplicate login pages, and some sites do not have a login page.
\jshelter{} is more accurate in fingerprinting detection when compared with
the scenario when FPMON and DFPM have low confidence in fingerprinting detection
(they score 1 point). \jshelter{} is slightly worse compared to the
scenario in which the other tools are confident that they detected
fingerprinting. The differently evaluated pages are typically borderline cases.
For example, \jshelter{} does not
detect fingerprinting on Google and Facebook login pages, while both FPMON
and DFPM detect fingerprinting. As the number of accessed APIs is not high and users would
likely turn FPD off for these pages, we do not intend to modify FPD heuristics.

\begin{table}[t]
   \centering
   \begin{tabular}{|l|l|c|c|}
     \hline
                 &                         &    Home pages & Login pages \\
     \hline
     \hline
     \jshelter{} & fingerprinting detected &  96 (98.0\,\%)& 77 (95.1\,\%) \\
     \hline
     FPMON       & red                     &  79 (80.6\,\%)& 66 (81.5\,\%) \\
                 & red or yellow           &  96 (98.0\,\%)& 80 (98.8\,\%) \\
     \hline
     DFPM        & two or more dangers     &  70 (71.4\,\%)& 66 (81.5\,\%) \\
                 & at least one danger     &  98 (100\,\%) & 81 (100\,\%) \\
     \hline
   \end{tabular}
   \caption{Tested tools accuracy based on the manual crawl of the top 100 web pages according to the Tranco list.}
   \label{tab:fpdstudy}
 \end{table} 

\subsection{Network Boundary Shield}

\subsubsection{Localhost scanning}

\begin{figure}[h!]
  \centering
    \includegraphics[width=0.45\textwidth]{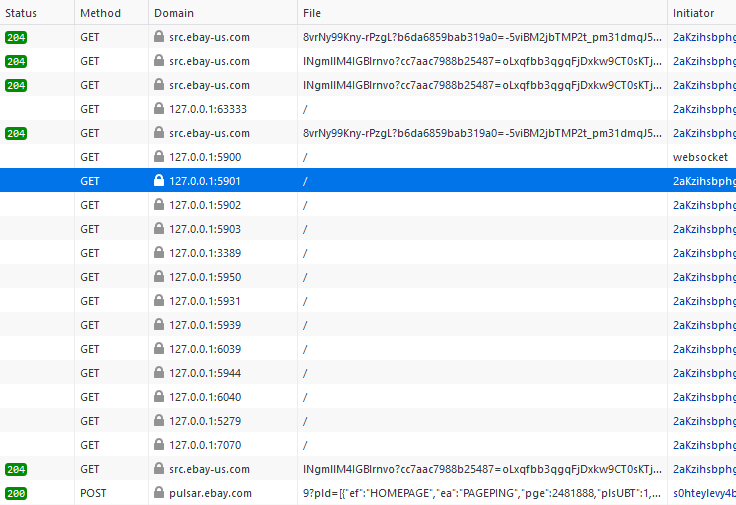}
  \caption{eBay web page scanning the local computer for open ports.}
 \label{fig:ebay-scan}
\end{figure}

Some web pages, like \url{ebay.com}, scan (some users) for open local TCP
ports to detect bots with open remote desktop access or possibly to create a
fingerprint. The web page instructs the browser to connect to the
\emph{localhost} (\verb|127.0.0.1|) and monitors the errors to detect if the
port is opened or closed. See Fig.~\ref{fig:ebay-scan} for an example.

When we developed NBS we did not anticipate localhost
port scanning. When we first encountered the eBay port scanning case, we knew
that this behaviour should trigger NBS as the requests cross network boundaries.
We accessed \url{ebay.com}, detected
the scanning by Web Developer Tools (Fig.~\ref{fig:ebay-scan}) and checked that
NBS is indeed triggered and works as expected.

\subsubsection{Comparison with Private Network Access}

Recently, Google announced Private Network Access
(PNA)\footnote{\url{https://developer.chrome.com/blog/private-network-access-prefilght/}}
that should become W3C
standard\footnote{\url{https://wicg.github.io/private-network-access/}}. PNA
solves the same problem as NBS, but the
solution is different. PNA-compatible browsers send HTTP Requests to the local
networks with the additional header: \\ \verb|Access-Control-Request-Private-Network: true|.

The local resource can allow such access with HTTP reply header
\verb|Access-Control-Allow-Private-Network:| \verb|true|. If it does not, the browser
blocks the access.

NBS works differently. Firefox version leverages DNS API to learn that a public
web page tries to access the local network and blocks the request before the
browser sends any data. Chromium-based browsers do not support DNS API, so the
first request goes through. NBS learns the IP address during the reply
processing. NBS blocks any future request before it is made once it learns the
IP address during the reply processing. Hence, NBS limits the network bandwidth
and prevents any state modification on a local node that may be caused
by a request going through, except for the learning phase in
Chromium-based browsers. Both approaches solve threat T3; it is
up to the user what solution they prefer.

Note that Google plans to fully deploy Chrome PNA in version 113, so Chrome
users without \jshelter{} or another webextension with similar capabilities are not
protected at the time of the writing of this paper\footnote{\url{https://chromestatus.com/feature/5737414355058688}}.

\subsection{Feedback from users}

\jshelter{} is available in \url{addons.mozilla.org}, Chrome
Store, and Opera Store from the early development stages. We employ the release early,
release often strategy, but we do not release early if we are concerned about
possible security bugs in the new version.

\begin{figure}[h!]
  \centering
    \includegraphics[width=0.48\textwidth]{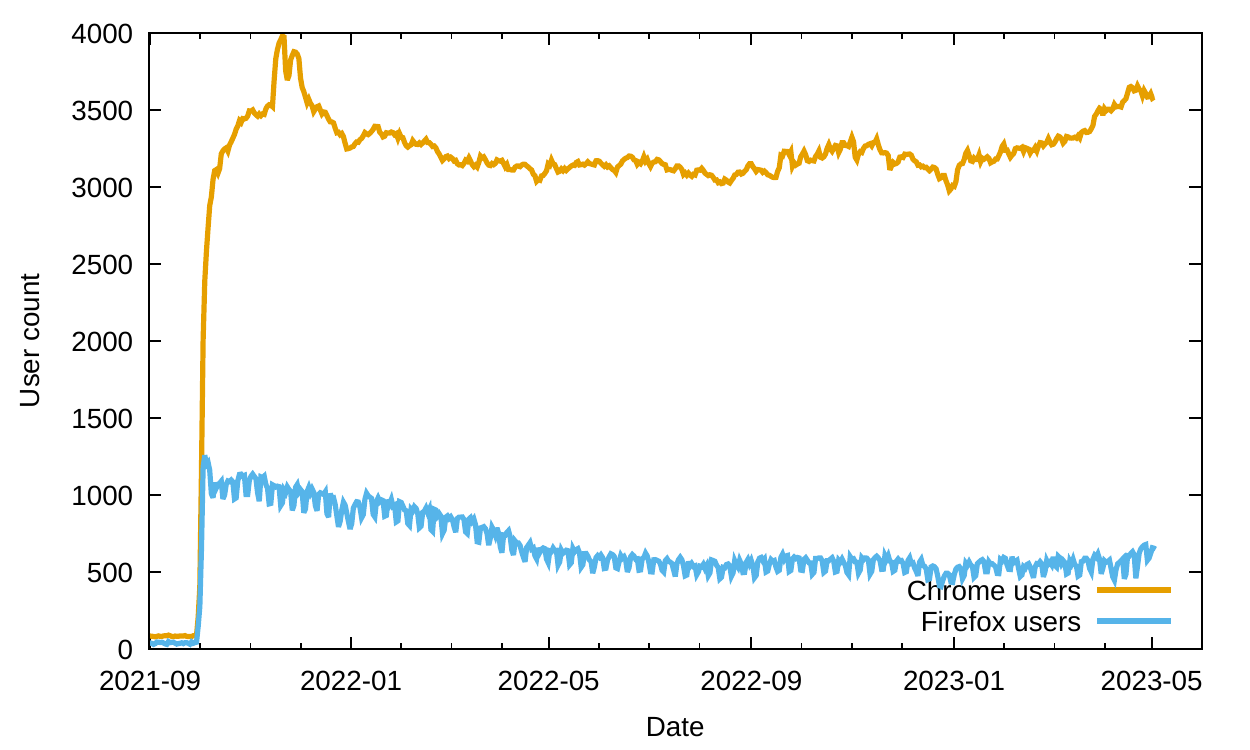}
  \caption{Number of \jshelter{} users.}
 \label{fig:users}
\end{figure}

Some users found \jshelter{} immediately after initial upload to webextension
sites. Nevertheless, the number of users increased massively only after an
announcement by \fsf{}. Figure~\ref{fig:users}, shows \jshelter{} users in time
in Firefox and Chrome. The graph shows that \jshelter{} has an
audience and users want to control their browsers. However, many users stopped
using the extension after a trial period. The decline in 2021 and 2022 seems to
be reverted and \jshelter{} gains users. The trend is clear for Chrome users,
the number of new Firefox users is smaller.

Based on the feedback from users, one of the reasons for the decline was that they
encountered too many broken or slow pages.
After we focused on fixing broken pages the decline
slowed down.

Another reason is that users do not understand the little lies fingerprint
prevention. They want to hide in the crowd (see Sect.~\ref{sec:3antifp}). The
most controversial protection is WebGL vendor, unmasked vendor, renderer, and unmasked
renderer spoofing. We do not know any list of real-world strings, and even if we
knew, we are not sure if we could avoid inconsistencies. Hence we decided that
the threat model defending from a fingerprinter not focused on revealing \jshelter{} users
allows for the generation of random strings per origin and session for the little
lies JSS profile (see Sect.~\ref{sec:jss}). Some users do not understand the
explanation even though we highlight that similar randomly generated strings are
already available through \verb|MediaDevices.prototype.enumerateDevices|, the
created profile is unique by design.

A common problem is that users do not understand what \jshelter{} is doing and
that several modules work in parallel and can be enabled and
configured separately. We tweaked the UI several times to make the UI as
straightforward as possible and we added explanations and want to add even more
explanations (for example, to the popup window).

\jshelter{} users also reported false-positive NBS detections when using
DNS-based filtering programs. Some DNS-based filters return the localhost IP address
for any blocked domain. In that case, NBS correctly detects that a public page
tries to access a local resource, blocks the request, and notifies the user.
We suggest that users reconfigure
their DNS
blocker to return 0.0.0.0 (invalid address). We also added options to turn
notifications completely off as
some users do not want to be notified at all.
We limited the number of notifications if they are
enabled and the web page accesses local nodes during a short time frame.
Additionally, we added explanation texts. Users do not report issues
with NBS notifications anymore.

\begin{figure}[h!]
  \centering
    \includegraphics[width=0.2\textwidth]{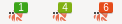}
  \caption{Interactive badge icons.}
 \label{fig:colors}
\end{figure}

Many webextensions report the number of blocked elements in the
badge icon. Previous research projects like Chrome Zero depicts currently
applied protections. Early \jshelter{} versions reported applied level as well,
but the feedback preferred showing the number of blocked elements and using
colours. We decided to (1) report the number of accessed API groups and (2)
report the likelihood of fingerprinting as a colour starting from shades of
green through yellow to shades of red. Figure~\ref{fig:colors} shows examples of
badged icons that received positive reception.

\section{Conclusion}
\label{sec:conclusion}

Previous research established that browser security, privacy, and customizability are
important
topics~\cite{fingerprinting_survey,js0,jucs15,clock_still_ticking,session-replay,block_me_if_you_can,gruss_rowhammer_js,forcepoint}.
The imminent danger of third-party cookie removal forces
trackers to employ even more privacy-invading techniques. Real-time bidding
leaves users easy targets for various attacks, including gaining information
about other applications running on the local computer~\cite{primeprobe}.
Moreover, continuous additions of new JavaScript APIs open new ways for fingerprinting
the browsers and gaining additional knowledge about the browser or user preferences
and physical environment.
One of the
major concerns is a need for more effective tools that everyday user wants to use. Current
methods to tackle web threats are list-based blockers that might be evaded with
a change of URL, specialised browsers, or research-only projects that are
quickly abandoned.

In contrast, \jshelter{} is a webextension that can be installed on major
browsers and does not require the user to change the browser and
routines. We integrate and improve several previous research projects like Chrome
Zero~\cite{js0}, little-lies-based fingerprinting prevention
\cite{PriVaricator,FPRandom}, and ideas for limiting APIs brought by Web API
Manager~\cite{webapi-vibrate}. \jshelter{} comes with a heuristic-based fingerprint detector
and prevents web pages from misusing the browser as a proxy to access the local network
and computer.
We solved issues with reliable environment modifications that stem
from insufficient webextension APIs that open many loopholes that previous research
exploited~\cite{PrimeProbe1JS0}. Nevertheless, at the time of the writing of
this paper, \jshelter{} does not allow to use Web Workers, which breaks some
pages. Besides \jshelter{}, we introduced
\NSCL{}.
Both \noscript{} and \jshelter{} include \NSCL{}. Moreover, \NSCL{} is available for other privacy-
and security-related webextensions.

In cooperation with \fsf{}, we
aim for long-term \jshelter{} development; thus, users' privacy and security
should be improved in the future.

\section{Acknowledgement}

This project was funded through the NGI0 PET Fund, a fund established by NLnet with financial support from the European Commission's Next Generation Internet programme, under the aegis of DG Communications Networks, Content and Technology under grant agreement No 825310 as JavaScript Restrictor and JShelter projects.
This work was supported in part by the Brno University of
Technology grant FIT-S-20-6293 and FIT-S-23-8209.
We thank Martin Bednář for capturing Figure~\ref{fig:ebay-scan}.

\bibliographystyle{ACM-Reference-Format}
\bibliography{biblio}

\end{document}